\documentclass[epj]{svjour}
\usepackage{graphics}
\begin{document}
\title{Novel Boron-10-based detectors for Neutron Scattering Science}
\subtitle{Helium-3-free detectors for large and small area applications: the Multi-Grid and the Multi-Blade prototypes}
\titlerunning{Novel Boron-10-based detectors for Neutron Scattering Science}
\authorrunning{F. Piscitelli}
\author{Francesco Piscitelli on behalf of the ILL/ESS/LiU collaboration for the development of the B10 detector technology in the framework of the CRISP project\thanks{\emph{Corresponding author:} francesco.piscitelli@esss.se}\inst{1,2,3}}     
%\mail {hcorrespondence authori}
%
\institute{Institut Laue-Langevin (ILL), 6, Jules Horowitz, 38042 Grenoble, France. \and European Spallation Source (ESS AB), P.O. Box 176, SE-221 00 Lund, Sweden. \and Department of Physics, University of Perugia, Piazza Universit\`a 1, 06123 Perugia, Italy.}
\date{Received: date / Revised version: date}
\abstract{
Nowadays neutron scattering science is increasing its instrumental power. Most of the neutron sources in the world are pushing the development of their technologies to be more performing. The neutron scattering development is also pushed by the European Spallation Source (ESS) in Sweden, a neutron facility which has just started construction. Concerning small area detectors ($\sim 1\,m^2$), the $\mathrm{^3He}$ technology, which is today cutting edge, is reaching fundamental limits in its development. Counting rate capability, spatial resolution and cost-effectiveness, are only a few examples of the features that must be improved to fulfill the new requirements. 
\\ On the other hand, $\mathrm{^3He}$ technology could still satisfy the detector requirements for large area applications ($\sim 50\,m^2$), however, because of the present $\mathrm{^3He}$ shortage that the world is experiencing, this is not practical anymore.
\\ The recent detector advances (the Multi-Grid and the Multi-Blade prototypes) developed in the framework of the collaboration between the Institut Laue-Langevin (ILL) and ESS are presented in this manuscript. In particular two novel $\mathrm{^{10}B}$-based detectors are described; one for large area applications (the Multi-Grid prototype) and one for application in neutron reflectometry (small area applications, the Multi-Blade prototype). 
\PACS{
     {29.40.-n}{radiation detectors}   %\and
  %    {PACS-key}{discribing text of that key}
     } % end of PACS codes
} %end of abstract
\maketitle
\section{Introduction}\label{intro}
\begin{figure}[!ht]
\centering
\resizebox{0.5\textwidth}{!}{\includegraphics{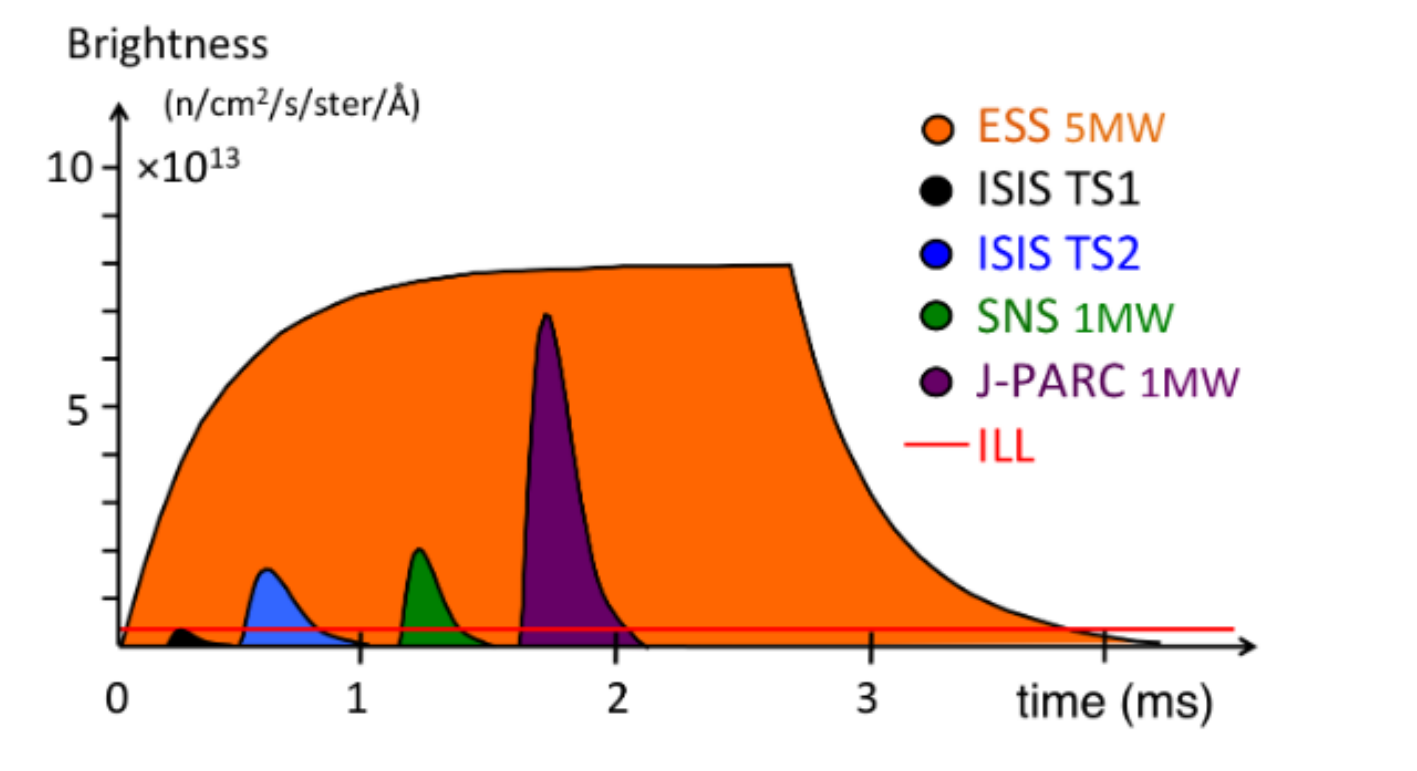}}
\caption{\footnotesize The ESS long pulse compared to the pulses of existing spallation sources in the world and to the steady flux available at reactor sources~\cite{esstdr}.}\label{puls}
\end{figure}

Neutron scattering science is increasing its instrumental power and consequently improved performance are requested to the detection systems. In particular the upcoming neutron facility ESS will need a first suite of instruments fully operational already in 2019~\cite{esstdr}. The peak brightness at the ESS will be higher than that of any of the existing short pulse sources and it will be one order of magnitude higher than that of the world's leading continuous source. The time-integrated brightness at ESS will also be one or two orders of magnitude larger than is the case at leading pulsed sources today. ESS will be a long pulse source, with an average beam power of $5\,MW$ delivered to the target station, fig.~\ref{puls} shows the expected pulse of ESS compared to the pulses of existing spallation sources in the world and to the steady flux available at reactor sources. 
\\ $\mathrm{^{3}He}$ gaseous detectors and scintillating detectors are the two principal technologies employed in thermal and cold neutron detection in all the facilities at present for any kind of intrument. These technologies are already close to the cutting-edge development which is sufficient to assure the performance of the instruments in neutron facilities at present. The needs of future neutron scattering science, including those of ESS, can not be met with today technologies if further studies and developments will not be performed~\cite{gebauer1,cooper,rhwessdet}.
\\ Because of the shortage of $\mathrm{^3He}$~\cite{shea,kouzes3,zeitsear}, it is crucial to find an alternative, efficient and cost-effective way to detect thermal and cold neutrons for large area applications ($\sim 50\,m^2$), e.g. chopper spectrometers like IN5~\cite{in5oliv} at ILL.
\\ Even though a limited quantity of $\mathrm{^3He}$ would be available for small area detectors ($\sim 1\,m^2$), the requirements for the future detectors can not be fulfilled by this technology. The main goal to be achieved for small area applications is to expand the detector performance mostly in terms of counting rate capability and spatial resolution, as well as cost-effectiveness. The development of small area detectors is focused on reflectometry applications because is where the requirements are more challenging. 
\\ In tab.~\ref{tab1} are summarized the features of the state of the art detector technology at existing facilities for chopper spectrometers and reflectometers.
\begin{table}[!ht]
\begin{center}
\footnotesize{
\begin{tabular}{|l l||c|c|c|c|c|c|}
  \hline
  \hline
 Instrument & Facility  & active area     & spatial res.       & efficiency  & global rate  & local rate \\
            &           &  ($m \times m$) &  ($mm \times mm$)  &             & ($s^{-1}$)    & ($s^{-1}cm^{-2}$)\\
 \hline
 \hline
chopper spectrometers  & & & & & & \\
\hline
IN5~\cite{in5oliv} & ILL & $10 \times 3$  & $ 26 \times 26$     & $\sim74\% \,@\, 1.8$\AA & - & - \\
\hline
\hline
reflectometers  & & & & & & \\
 \hline
FIGARO~\cite{figaro}  & ILL   & $0.5 \times 0.25$  & $ 2 \times 7.5$ & $\sim63\% \,@\, 2.5$\AA & $3\cdot10^7$ & $23\cdot10^3$  \\
\hline
INTER~\cite{inter}    & ISIS  & $0.2 \times 0.2$   & $ 1 \times 1$   & - & - & - \\
 \hline
 \hline
\end{tabular}}
\caption{\footnotesize Detectors features on instruments at existing facilities, where figures are publically available.}
\label{tab1}
\end{center}
\end{table}
\\ Table~\ref{tab2} summarizes the requirements at ESS for large area detectors for chopper spectrometers and for small area detector for reflectometry applications. The flux at sample must be also interpreted as the maximum rate the detector should tolerate. It is not unlikely that the full beam is completely reflected toward the detector in a neutron reflectometry experiment. 
\begin{table}[!ht]
\begin{center}
\footnotesize{
\begin{tabular}{|l|c|c|c|c|c|}
  \hline
  \hline
 ESS instrument type & active area  & wavelength band & spatial resolution   & time resolution & max flux at sample \\
                     &   ($m^2$)    &         (\AA)   &    ($mm \times mm$)  &     ($\mu s$)   & ($s^{-1}cm^{-2}$) \\
 \hline
 \hline
chopper spectrometers~\cite{trexprop} & $50$         & $[0.8,20]$      &$20 \times 20$        & $10$            & $10^7$ \\
 \hline 
 reflectometers~\cite{freiaprop}      & $\sim 1$         & $[2,23]$        & $0.5 \times 2$       & $100$       & $10^9$ \\
 \hline
 \hline
\end{tabular}}
\caption{\footnotesize Summary of generalized detector requirements for chopper spectrometers and refletometers at ESS.}
\label{tab2}
\end{center}
\end{table}
\\ The Multi-Grid~\cite{jonisorma,mgpat} prototype has been developed in the framework of the collaboration between ILL and ESS in order to address the problem of the $\mathrm{^3He}$ shortage for large area detectors. It is a gaseous detector that contains $30$ $\mathrm{^{10}B_4C}$ layers as neutron converters. Each neutron can be converted in one of these layers that are crossed orthogonally by the neutron beam. Several prototypes using the Multi-Grid design have been built and tested, showing the feasibility of such a design for large area coverage with a suitable neutron detection efficiency.  
\\ On the other hand, the Multi-Blade~\cite{buff3,framb} prototype, already introduced at ILL in 2005~\cite{buff1}, is a small area detector which wants to push the limit of spatial resolution beyond that of $\mathrm{^3He}$-based detectors for high flux applications. In particular it has been conceived to be suitable for neutron reflectometry instruments, though potential applications could be broader than that. A detailed study on this prototype has been carried out at ILL showing the improvements in terms of detector features of such a design. 
\\ The Multi-Grid and the Multi-Blade prototypes are described and the results are illustrated in the following sections. 

\section{Large area applications: the Multi-Grid prototype}\label{mgsect}
\begin{figure}[!ht]
\centering
\resizebox{0.7\textwidth}{!}{\includegraphics{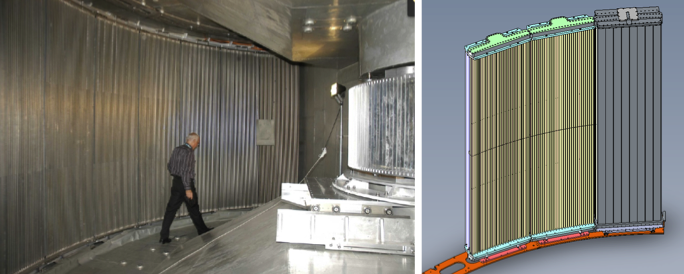}}
\caption{\footnotesize A picture of the vacuum chamber of IN5 at ILL equipped with $\mathrm{^3He}$-tubes (left). A drawing of the Multi-Grid $3\times0.8\,m^2$ area demonstrator nearby the $\mathrm{^3He}$-detectors of the instrument (right).}\label{figmg3}
\end{figure}

\begin{figure}[!ht]
\centering
\resizebox{0.45\textwidth}{!}{\includegraphics{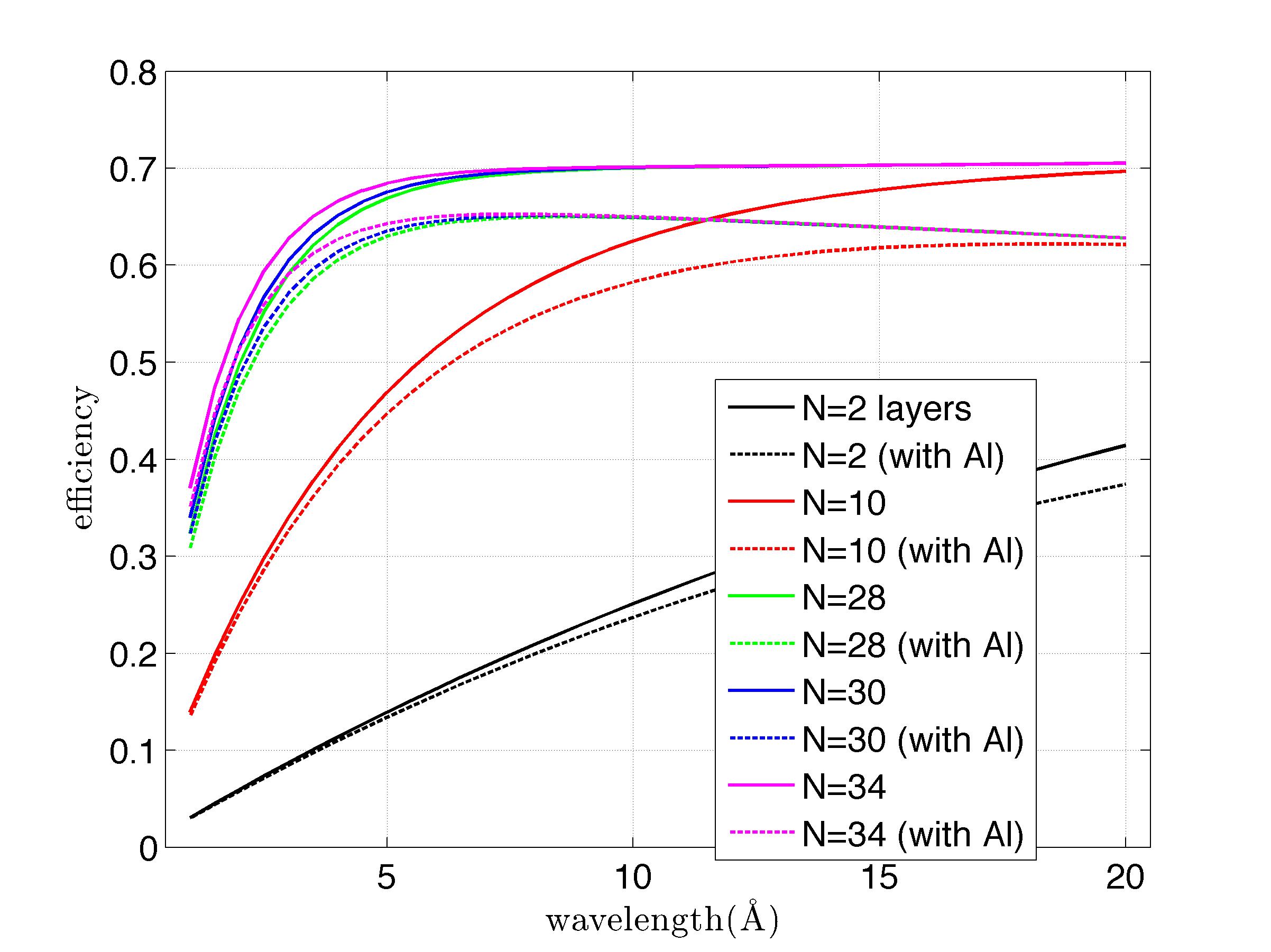}}
\resizebox{0.45\textwidth}{!}{\includegraphics{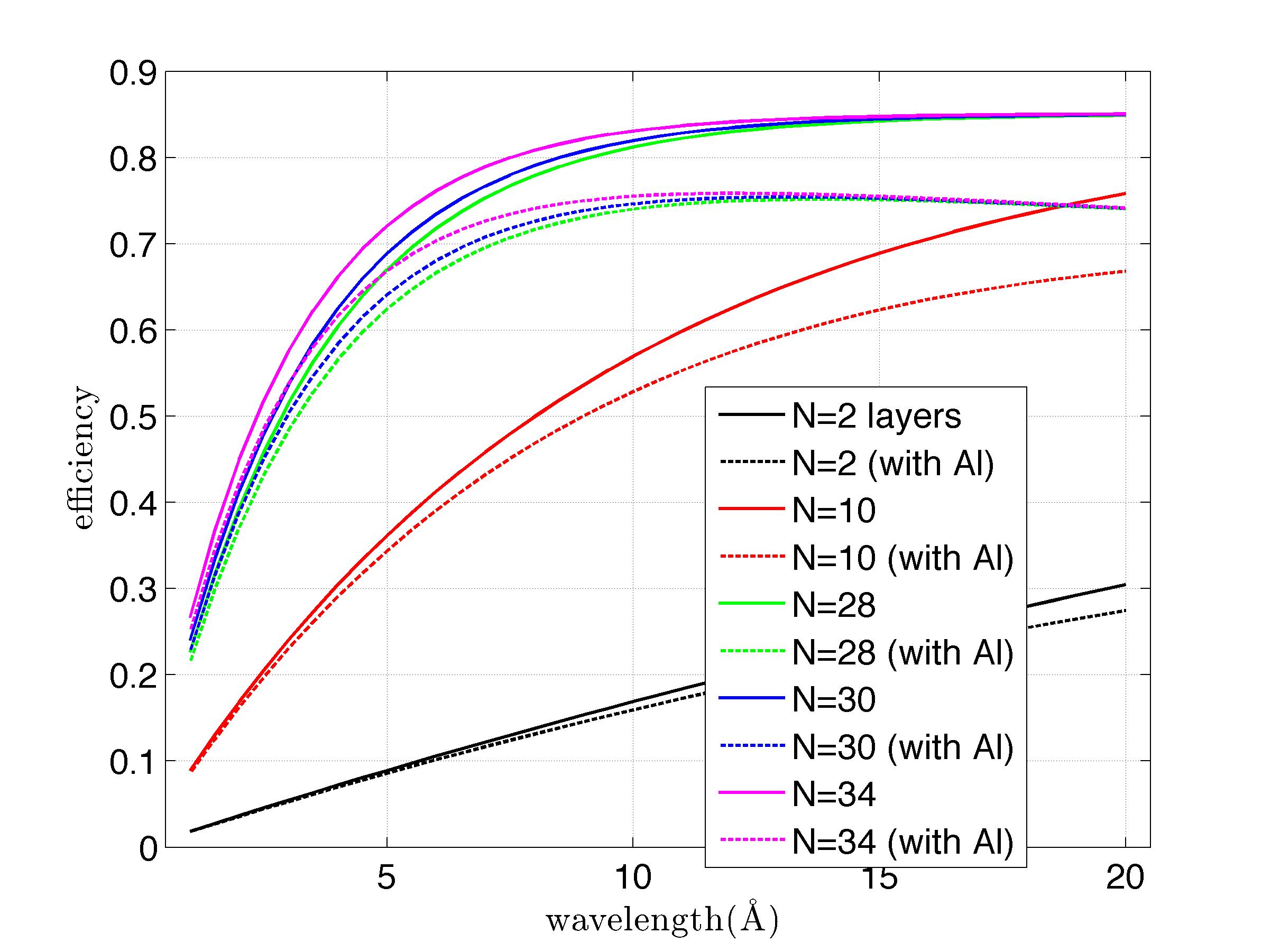}}
\caption{\footnotesize Expected efficiency as a function of the neutron wavelength for a detector made up of $2,10,28,30$ and $34$ $\mathrm{^{10}B_4C}$-layers. The converter thickness is $1\,\mu m$ for all the layers in the plot on the left and $0.5\,\mu m$ for the plot on the right. The contribution of the Al-scattering from substrates and detector window is also taken into account.}\label{figmgeff}
\end{figure}

\begin{figure}[!ht]
\centering
\resizebox{0.7\textwidth}{!}{\includegraphics{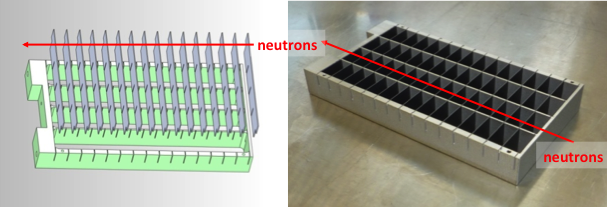}}
\caption{\footnotesize A drawing of a frame where $\mathrm{^{10}B_4C}$-coated blades are inserted (left) and fully assembled grid (frame and blades) of the prototype (right).}\label{figmg1}
\end{figure}

\begin{figure}[!ht]
\centering
\resizebox{0.7\textwidth}{!}{\includegraphics{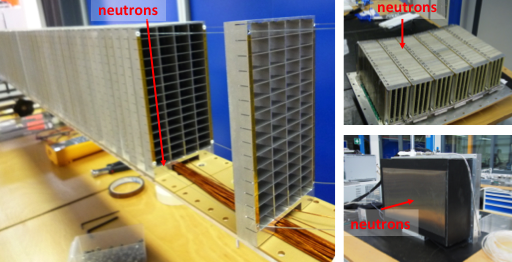}}
\caption{\footnotesize A $96$-grid prototype of active area $0.08\times2\,m^2$ is shown on the left, during its contruction. A $0.3\times0.5\,m^2$ prototype is shown on the right.}\label{figmg2}
\end{figure}

Several version of the Multi-Grid detector have been implemented at ILL in collaboration with ESS in the framework of the CRISP project (http://www.crisp-fp7.eu/). With the construction of these prototypes the reliability and the actual performance of this design have been shown. A new prototype of $3\times0.8\,m^2$ active area is being built to demonstrate the feasibility of a large area detector for neutron chopper spectrometers, e.g. IN5 at ILL~\cite{in5oliv}. The IN5 Time-of-Flight spectrometer is used as a benchmark for performance and geometry requirements. IN5 was chosen for its size of about $\sim 30\,m^2$. It is the largest chopper spectrometer at ILL, finding a replacement for $\mathrm{^3He}$-based detectors over such a surface is the main goal to validate the feasibility of large areas detectors based on $\mathrm{^{10}B_4C}$. The Multi-Grid design must be able to cover a sensitive area above $30\,m^2$ with about $2\times2\,cm^2$ spatial resolution. Figure~\ref{figmg3} shows a picture of the vacuum chamber of IN5 equipped with $12\times32$ $\mathrm{^3He}$-tubes and a drawing of the $3\times0.8\,m^2$ area demonstrator nearby the $\mathrm{^3He}$ detectors of the actual instrument. 
\\ A Multi-Grid is a proportional gaseous detector in which the neutron conversion into charged particles is obtained by using solid $\mathrm{^{10}B_4C}$-layers. $\mathrm{^{10}B}$ was chosen as the neutron converter because of its large absorption cross-section for thermal neutrons ($\sigma_{abs}=3838b$), and relatively high energy of its reaction products~\cite{jonisorma}. In the actual prototype the $\mathrm{^{10}B_4C}$ was chosen due to its conductivity, thermal and chemical stability compared to pure $\mathrm{^{10}B}$. Due to the limited efficiency that can be achieved with a single layer~\cite{gregor} (if it is not operated under a grazing angle as it is in the Multi-Blade concept~\cite{framb}), many layers are necessary to reach a suitable detection efficiency. If a single layer is used it must be operated a grazing angle because the neutron absorption path in the layer travels closer to the surface. Consequently the neutron capture fragments have more chance to escape the layer and thus to contribute to the efficiency. At a normal incidence a single layer is only about $5\%$ efficient at thermal energies. 
\\ An optimization of the layer thicknesses, the number of layers necessary, and considerations on the substrate effects are required as well as a minute mechanical study. According to~\cite{fratheo}, in order to get a detection efficiency around $50\%$ while keeping the mechanics reasonably simple, a good choice for the number of layers is around $30$. The optimal thickness of the layers should be fixed at $1\,\mu m$ if we consider a monochromatic neutron beam at $2.5$\AA~\cite{fratheo}. The interesting neuron wavelength range for the scientific case of chopper spectrometer instruments is about from $1$\AA\, to $20$\AA\, which corresponds to energies from $82\,meV$ to $200\,\mu eV$. Further optimization is possible if there are requirements on a particular neutron wavelength distribution or efficiency. It has been shown in~\cite{fratheo} that a reasonable optimization is the one that considers the barycenter of the wanted neutron wavelength distribution, i.e. the $2.5$\AA\, will provide a detector optimized for short wavelengths. Figure~\ref{figmgeff} shows the expected efficiency as a function of wavelength for a detector made up of $2,10,28,30$ and $34$ $\mathrm{^{10}B_4C}$-layers, in the two cases in which all the layers are $1\,\mu m$ or $0.5\,\mu m$ thick. The detector with $0.5\,\mu m$ layers is optmized for long wavelengths. The contribution given by the scattering or absorption in the aluminium substrates ($0.5\,mm$ thick) is also taken into accounts in the plots in fig.~\ref{figmgeff}, as well as a $2.5\,mm$ entrance Al-window. The decreasing efficiency at longer wavelengths is due to the absorption of neutrons by the detector entrance Al-window. No decrease can be observed if the entrance window is neglected. 
\\ Several prototypes made up of $28$ and $30$ layers have been built and tested. A $34$-layer detector of large area is under construction. 
\\ A Multi-Grid detector is composed of a gas vessel filled with grids electrically insulated one from another and stacked to make square or rectangular counters~\cite{mgpat}, i.e. tubes. Each grid is made up of a frame in which blades, coated with $\mathrm{^{10}B_4C}$~\cite{carina} on both sides, are inserted (see fig.~\ref{figmg1}). Both the frame and the blades are made of aluminium which has a low scattering and absorption cross-section for neutrons~\cite{bruproceed}.
\begin{figure}[!ht]
\centering
\resizebox{0.5\textwidth}{!}{\includegraphics{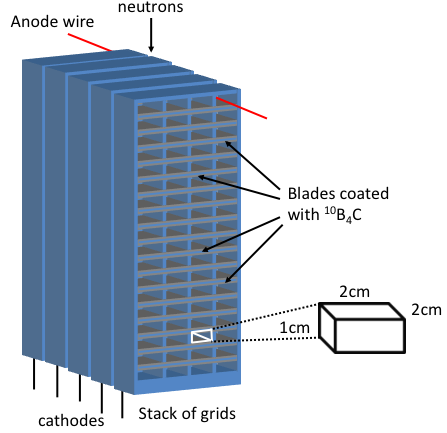}}
\caption{\footnotesize The grids are stacked to form rectangular tubes where an anode wires are inserted.}\label{schemMG}
\end{figure}
\\ The grids, stacked one after the other, act as a segmented cathode (see fig(s).~\ref{schemMG} and~\ref{figmg2}). The readout is performed by the cathode grids and anode wires. Each of these anode wires is placed in the middle of a tube formed by the stacked grids. Individual and charge division readouts have both been implemented. The grids are electrically insulated and each of them is connected to a charge amplifier. Each anode wire is either connected directly to a charge amplifier (individual readout) or is connected to the other wires through resistors. This allows to readout many wires with only two charge amplifiers connected at the ends of the resistive chain. The position of the neutron interaction point is reconstructed by the coincidence between grid and wire signals.
\\ Figure~\ref{figmg2} shows two of the assembled Multi-Grid prototypes. Both of them used $28$ converter layers. The first, with its $96$ grids, had an active area of $0.08\times2\,m^2$~\cite{bruproceed,buff2,khaplanov}, and the second demonstrator of about $0.3\times0.5\,m^2$. The second prototype~\cite{in6procc} was installed on the Time-of-Flight chopper spectrometer IN6~\cite{in6papold} at ILL placed side by side to the conventional $\mathrm{^3He}$ detectors of the instrument. 
\\ The detector was tested with both $\mathrm{Ar/CO_2}$ ($90:10$) gas mixture and $\mathrm{CF_4}$~\cite{bruproceed}. Counting curves and Pulse Height Spectra (PHS) have been measured for several different stopping gas pressures in order to confirm that the detector can be operated at either atmospheric pressure or even below. 
\\ The neutron detection efficiency was measured on our test beam line at ILL, CT2, with a monochromatic neutron beam of $2.5$\AA. As shown in~\cite{jonisorma}, an $\mathrm{^3He}$-multi-tube detector was used as a reference and an efficiency of $(46.8\pm0.3)\%$ was found for the detector with $28$ converter layers. The position resolution was also measured and it resulted to be $2\times2\,cm^2$. This resolution is given by the size of the voxel identified by a single grid and a wire (see fig.~\ref{schemMG}). It has been proved that a finer spatial resolution can be obtained by calculating the center of gravity of the induced charge on adjacent grids. 
\\ The $96$-grids-detector was scanned in order to quantify the uniformity in the efficiency. It has been already proven~\cite{carina}, that the $\mathrm{^{10}B_4C}$ coatings show about a $13\%$ variation in thickness due to the deposition method, this affects the detector uniformity in the range of $1\%$~\cite{jonisorma}. 
\\ An accurate study on the response of the Multi-Grid detector in presence of $\gamma$-ray backgorund is reported in~\cite{khap}. A high level of discrimination between neutron and photon signals can be reached with $\mathrm{^3He}$ detectors. The sensitivity of a neutron detector to $\gamma$-rays is a very important characteristic, as it defines the best achievable signal-to-noise ratio. In a neutron facility the $\gamma$-ray background is produced by the interaction of the neutrons with the parts of the instrument. Neutron guides, collimators, shielding, beam stoppers and choppers, are only a few of the objects where a neutron can be absorbed giving rise to $\gamma$-ray background. The majority of efficient neutron absorbers, in particular $\mathrm{^{10}B}$, $\mathrm{^{113}Cd}$, $\mathrm{^{115}Cd}$ and $\mathrm{^{157}Gd}$, emit one or more $\gamma$-rays for each neutron captured. The energy of these $\gamma$-rays span a very wide range that goes from a few tens of $keV$ up to about $10\,MeV$. Considering that the scattering cross sections of many samples tend to be relatively low, the neutron intensity that carries useful information can be many orders of magnitude lower than the flux of $\gamma$-rays~\cite{khap}.
\\ The physical effects and geometric considerations that affect the sensitivity to $\gamma$-rays in gas-based detectors for thermal neutrons have been deeply investigated. It has been found that the $\gamma$-ray rejection do not need to be lower in $\mathrm{^{10}B}$-based detectors than in $\mathrm{^3He}$ tubes. A correctly-chosen energy threshold and operating voltage allow for an equally high $\gamma$-ray rejection as in a $\mathrm{^3He}$ detector. In particular for the Multi-Grid detector, it has been shown that sensitivities below $10^{-6}$ can be reached by setting an energy threshold to $100\,keV$ and operating the Multi-Grid at $900\,V$. At this voltage the equivalent gas gain is below $100$, consequently $\gamma$-rays can be discriminated by their energy and only loosing a few percent in the neutron detection efficiency.
\\ With the Multi-Grid demonstrator, installed on IN6 in order to be tested in a real instrument environment, it has been shown the suitable Time-of-Flight resolution of the $\mathrm{^{10}B}$-based detector as well as its better solid angle coverage with respect to $\mathrm{^3He}$ tubes. Considering the dead spaces, i.e. solid angle coverage of the Multi-Grid detector, this is only $3\%$ less efficient than the IN6 detection system at $4.6$\AA \,and $9\%$ more efficient at $4.1$\AA~\cite{in6procc}. 
\\ During the tests on IN6 it became clear that not only $\gamma$-ray and neutron background was present: the detector showed a count rate that could not be correlated neither to $\gamma$-ray or neutron events. This background was not modulated when the beam was pulsed and was present irrespective of the instrument's shutter state or even the reactor operation. The found the source of such background are the naturally-occurring concentrations of uranium and thorium in aluminium~\cite{Al1}. The concentration of radioisotopes in materials~\cite{Al2} is a well known topic in the the chemical industry and it has also been investigated in the field of the neutron detectors~\cite{Al3}. Two solutions are currently being investigated to eliminate this background. One is to deposit a thin nickel layer at the surface of aluminium in order stop the alpha particles before they reach the gas~\cite{prax}. The other solution is to use $\mathrm{U/Th}$-free aluminium for which the concentration of uranium and thorium is $2-3$ orders of magnitude lower than in standard aluminium alloys~\cite{hyd}. Both this solutions reduce the background of several order of magnitudes. Respective costs will likely be the deciding factor between these two approaches.
\\ The following step in this project is the construction of the $3\times0.8\,m^2$ large area demonstrator. This detector will demonstrate the feasibility and cost of large-scale production of both the mechanical parts as well as the $\mathrm{^{10}B_4C}$-layers required~\cite{in6procc}.

\section{Small area applications: the Multi-Blade prototype}\label{mbsect}
\begin{figure}[!ht]
\centering
\resizebox{0.4\textwidth}{!}{\includegraphics{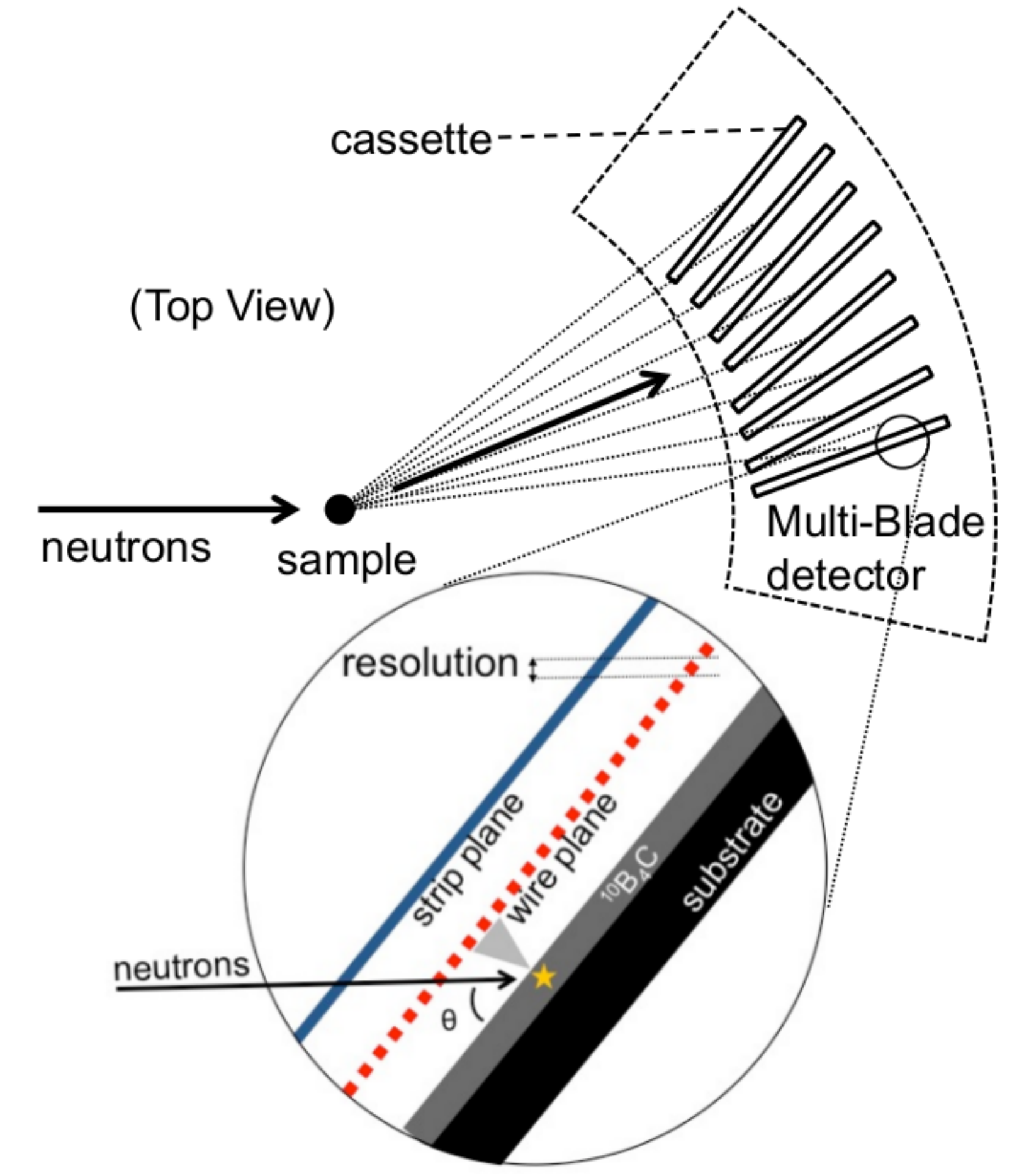}}
\caption{\footnotesize A drawing of the Multi-Blade detector, several identical units (cassettes) are aligned to form a detector. Each module is an independent MWPC.}\label{schemMB}
\end{figure}

\begin{figure}[!ht]
\centering
\resizebox{0.7\textwidth}{!}{\includegraphics{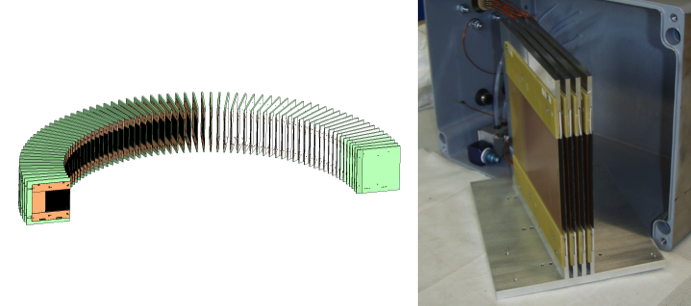}}
\caption{\footnotesize Schematic of the Multi-Blade detector made up of several independent and identical units called \emph{cassettes} (left). A picture of the Multi-Blade prototype composed of $4$ cassettes (right).}\label{figmb}
\end{figure}

The Multi-Blade prototype~\cite{framb} is a small area detector for neutron reflectometry applications~\cite{figaro}. Although the amount of $\mathrm{^3He}$ needed for a small area detector will be available in the future, the Multi-Blade wants to push the limit of $\mathrm{^3He}$-based detectors in terms of spatial resolution and counting rate capability. The Multi-Blade is a Multi Wire Proportional Chamber (MWPC) operated at atmospheric pressure (see fig.~\ref{schemMB}). This detector uses $\mathrm{^{10}B_4C}$ converters at grazing angle with respect to the incoming neutron beam. The angled geometry improves both its spatial resolution and its counting rate capability. The use of the $\mathrm{^{10}B_4C}$ conversion layer at grazing angle also increases the detection efficiency.
\\ The Multi-Blade prototype is conceived to be modular in order to be adaptable to different applications. A significant concern in a modular design is the uniformity of the detector response. Several effects might contribute to degrade the uniformity and they have to be taken into account in the detector concept: overlap between different substrates, coating uniformity, substrate flatness and parallax errors. 
\\ Each module of the Multi-Blade, a so-called \emph{cassette}, is an independent MWPC equipped with the neutron converter and a two-dimensional readout system. Wires are placed orthogonally to cathode strips (see fig.~\ref{schemMB}); the interaction point of the neutron is reconstructed by the coincidence wires and strips. 
\\ The fully assembled detector is composed of several cassettes as shown in the schematic in fig.~\ref{figmb}. 
\\ Several approaches in the prototype design have been studied: number of converters, read-out system and materials to be used. Two versions of the Multi-Blade prototype have been built focusing on its different issues and features.
Their detection efficiency and uniformity were measured on the test beam line CT2 at ILL. It has been shown in~\cite{framb} that a suitable detection efficiency for reflectometry applications can be achieved with such a detector. The main issue in the Multi-Blade design is the uniformity over its active surface. The cassettes overlap to avoid dead zones, and, in the switching between one cassette to another, a loss in efficiency can occur. Further studies are necessary to optimize the detector uniformity. 
\\ The detector is operated at atmospheric pressure. This makes it suitable to be operated in vacuum chambers. Moreover, cost effective materials can be used inside the detector because outgassing is not an issue. The materials used in the detector can pour out molecules that give rise to radicals. The presence of radicals in the gas drastically affects the detector functionality and aging. Since the gas is flushed those molecules are continuously evacuated.   
\\ The spatial resolution was measured~\cite{framb} to be $0.275\times4\,mm^2$. The reasonable limit in the resolution that can be reached with $\mathrm{^3He}$ detectors is around $1\,mm$. In many areas of soft and hard matter research science, the amount of material to investigate is rather limited. Partly because the fabrication of larger samples is too expensive or not feasible, yet, partly because the interesting features depend on the size. The development of a neutron reflectometer optimized for small samples is under study~\cite{rainbow1,estiaprop}. There is a great interest in expanding the technique of neutron reflectometry beyond static structural measurements of layered structures to kinetic studies~\cite{cubitt2}. In order to perform these studies the neutron wavelength to position encoding is necessary~\cite{rainbow2}, but, due to the practical limits in the actual spatial resolution of $\mathrm{^3He}$-based detectors this concept is probably not practical. Therefore, the development of an area detector with $0.2\,mm$, required in one dimension only, is crucial~\cite{cubitt2}. The Multi-Blade can easily fulfill this requirement. The Multi-Blade concept is a promising alternative, to accomplish the high spatial resolution and the high counting rate capability which is in principle $10$ times higher than conventional $\mathrm{^3He}$-based detectors~\cite{framb}. 

\section{Conclusions}\label{conclu}
The Multi-Grid concept has been successfully introduced and tested. It has been shown, with the construction of several prototypes, that the performance of such a design fulfills the needs of the intruments that requires large area detectors, i.e. chopper spectrometers. A neutron detection efficiency of about $50\%$ at $2.5\,$\AA\, can be reached with $30$ $\mathrm{^{10}B_4C}$ layers~\cite{jonisorma}. This detector is operated at atmospheric pressure with a continuous gas flow. This makes this detector suitable to be operated in vaccum chambers as well as the cost-effectiveness of the materials that can be used. 
\\  A $0.3\times0.5\,m^2$ area demonstrator~\cite{in6procc} has been tested in the Time-of-Flight chopper spectrometer IN6 at ILL placed side by side to the conventional $\mathrm{^3He}$ detectors of the instrument. The suitable Time-of-Flight resolution of the $\mathrm{^{10}B}$-based detector as well as its better solid angle coverage with respect to $\mathrm{^3He}$ tubes have been shown. $\mathrm{U/Th}$-free aluminium and nickel plating of aluminium have been investigated as the two solutions to reduce the naural $\alpha$-emission of Al-contaminants. Backround is efficiently reduced and the respective costs will  be the deciding factor between these two approaches.
\\ Although the Multi-Grid detector efficiency is smaller than for $\mathrm{^3He}$ detectors, it can be considered suitable in the context of the $\mathrm{^3He}$ shortage. In addition to that, it has been shown that the wider solid angle coverage of the Multi-Grid design, i.e. its higher granularity, with respect to the $\mathrm{^3He}$ detectors, adequately compensates this smaller efficiency~\cite{in6procc}.
\\ An accurate study on the $\gamma$-ray sensitivity of such a technology has been performed to validate its reliability in terms of background $\gamma$-ray rejection~\cite{khap}.   
\\ A prototype of $3\times0.8\,m^2$ and exploiting $34$ layers is under construction. The expected efficiency is about $56\%$ for $2.5\,$\AA. This prototype will demonstate the feasibility of a large area detector that can be scaled up to a few tens of square meters. 
\\ In order to tackle the problem of the limitations of the $\mathrm{^3He}$ technology in terms of counting rate capability and spatial resolution for small area detector, the Multi-Blade prototype has been built and tested~\cite{framb}. In particular a high spatial resolution detector based on inclined $\mathrm{^{10}B_4C}$ layers suitable for neutron reflectometry instruments has been developed. This detector has been conceived to be modular in order to be more versatile: it is composed of modules called cassettes. All the issues that can arise from a modular design have been investigated. The detector is operated at atmospheric pressure. This makes it suitable to be operated in vacuum. Moreover, cost effective materials can be used inside the detector because outgassing is not an issue. It has been shown that a proper detection efficiency for reflectometry instruments can be achieved with such a detector. The presented Multi-Blade prototype showed a very high spatial resolution, it was measured to be about $0.3\times4\,mm^2$~\cite{framb}.
\\ Throughout the design, assembly and characterization of our $\mathrm{^{10}B}$-based prototypes, it has been shown that this technology is a viable replacement of the sparse $\mathrm{^3He}$, in terms of performance and cost. This  $\mathrm{^{10}B}$ detector technology will certainly form a central part of the landscape for future detectors for neutron scattering instruments. 

\begin{acknowledgement}
\textbf{\large Acknowledgements}
\medskip
\\ The work has been supported by the CRISP project (European Commission 7th Framework Programme Grant Agreement 283745) - http://www.crisp-fp7.eu/.
\end{acknowledgement}

\end{document}